\documentclass[final,leqno,onefignum,onetabnum]{siamltex1213}

\graphicspath{ {./figures/} }

\usepackage{caption}
\usepackage{subcaption}
\usepackage{verbatim}
\usepackage{listings}
\usepackage{algorithm}
\usepackage[noend]{algpseudocode}

\newcommand{\sectionname}{section}
\newcommand{\secref}[1]{\sectionname~\ref{#1}}
\newcommand{\figref}[1]{\figurename~\ref{#1}}

\usepackage{tikz}
\usepackage{pgfplots}
\usetikzlibrary{arrows,chains,positioning,fit,backgrounds,calc,shapes,
  shadows,scopes,decorations.markings,plotmarks}

\newcommand*{\tettextsize}{\footnotesize}
\tikzstyle{line} = [draw, -, thick]
\tikzstyle{nodraw} = [draw, fill, circle, minimum width=0pt, inner sep=0pt]
\tikzstyle{sieve} = [line, circle, font=\tettextsize, inner sep=0pt,
  minimum size=12pt]

\newif\ifmonochrome
\monochrometrue

\ifmonochrome
\tikzstyle{cell} = [sieve]
\tikzstyle{facet} = [sieve]
\tikzstyle{edge} = [sieve]
\tikzstyle{vertex} = [sieve]
\else
\tikzstyle{cell} = [sieve, fill=blue!60]
\tikzstyle{facet} = [sieve, fill=green!35]
\tikzstyle{edge} = [sieve, fill=red!35]
\tikzstyle{vertex} = [sieve, fill=blue!35]
\fi

\begin{document}

\title{Efficient mesh management in Firedrake using PETSc-DMPlex}

\author{
  Michael Lange\footnotemark[2]
  \thanks{Corresponding author; e-mail: \email{michael.lange@imperial.ac.uk}}
  \and Lawrence Mitchell\footnotemark[3]
  \and Matthew G. Knepley\footnotemark[4]
  \and Gerard J. Gorman\footnotemark[2]
}

\maketitle
\slugger{sisc}{xxxx}{xx}{x}{x--x}

\renewcommand{\thefootnote}{\fnsymbol{footnote}}
\footnotetext[2]{Department of Earth Science and Engineering, Imperial College London, UK}
\footnotetext[3]{Department of Mathematics and Department of Computing, Imperial College London, UK}
\footnotetext[4]{Computational and Applied Mathematics, Rice University, USA}
\renewcommand{\thefootnote}{\arabic{footnote}}

\begin{abstract}
The use of composable abstractions allows the application of new and
established algorithms to a wide range of problems while automatically
inheriting the benefits of well-known performance optimisations. This
work highlights the composition of the PETSc DMPlex domain topology
abstraction with the Firedrake automated finite element system to
create a PDE solving environment that combines expressiveness,
flexibility and high performance. We describe how Firedrake utilises
DMPlex to provide the indirection maps required for finite element
assembly, while supporting various mesh input formats and runtime
domain decomposition. In particular, we describe how DMPlex and its
accompanying data structures allow the generic creation of
user-defined discretisations, while utilising data layout
optimisations that improve cache coherency and ensure overlapped
communication during assembly computation.

\end{abstract}


\begin{AMS}\end{AMS}

\pagestyle{myheadings}
\thispagestyle{plain}
\markboth{M. Lange et al.}{DMPlex mesh management in Firedrake}

\section{Introduction}

The separation of model description from implementation facilitates
multi-layered software stacks consisting of highly specialised
components that allow performance optimisation to happen at multiple
levels, ranging from global data layout transformations to local kernel
optimisations. A key challenge in designing such multi-layered systems
is the choice of abstractions to employ, where a high degree of
specialisation needs to be complemented with the generality required
to facilitate the utilisation of third-party libraries and thus
promote code reuse. The use of high-level domain-specific languages
(DSL) and composable abstractions allows existing algorithms and
optimisations to be inserted into this hierarchical framework, and applied
to a much wider range of problems.

In this paper we describe the integration of the DMPlex mesh topology
abstraction provided by the PETSc library~\cite{Balay2014} with
Firedrake, a generalised system for the automation of the solution of
partial differential equations using the Finite Element method
(FEM)~\cite{Rathgeber2015}. We outline how DMPlex is utilised in
Firedrake to provide the required mapping between topological entities
and degrees of freedom (DoFs), while supporting various mesh input
formats, run-time domain decomposition and mesh renumbering
techniques. In particular, we describe how DMPlex and its accompanying
data structures allow the generic creation of user-defined
discretisations, while utilising data layout optimisations that
optimise cache coherency and ensure computation-communication overlap
during finite element assembly.

\section{Background}

\subsection{Firedrake}

Firedrake is a novel tool for the automated solution of Finite Element
problems defined in the Unified Form Language (UFL)~\cite{Alnaes2014},
a domain-specific language (DSL) for the specification of partial
differential equations in weak form pioneered by the FEniCS
project~\cite{Logg2012}. Firedrake imposes a clear separation of
concerns between the definition of the problem, the local
discretisation defining the computational kernel used to compute the
solution and the parallel execution of this kernel over a given data
set~\cite{Rathgeber2015}. These multiple layers of abstraction allow
various types of optimisation to be applied during the solution
process, ranging from high-level caching of mathematical forms to
compiler-level optimisations that leverage threading and vectorisation
intrinsics within the assembly kernels.

A key component to achieving performance in Firedrake is PyOP2, a
high-level framework that optimises the parallel execution of
numerical kernels over unstructured mesh data~\cite{Rathgeber2012}.
PyOP2 represents mesh entities as {\em sets} and connectivity between
them as {\em mappings}, where input data to the compiled kernel is
either accessed directly or indirectly via a {\em mapping}.
In parallel PyOP2 is able to overlap halo data communication with
kernel computation during the execution loop due to a specialised data
ordering within {\em sets}~\cite{Mudalidge2012}.

\subsection{DMPlex}

PETSc's ability to manage unstructured meshes is centred around
DMPlex, a data management object that encapsulates the topology of
unstructured grids and provides a wide range of common mesh management
functionalities to application programmers~\cite{Lange2015}. As such
DMPlex provides a domain topology abstraction that decouples user
applications from the implementation details of common mesh-related
utility tasks, such as file I/O, domain decomposition methods and
parallel load balancing~\cite{Knepley2015}, which increases
extensibility and improves interoperability between scientific
applications through librarization~\cite{Brown2015}.

\begin{figure}[ht]
  \centering
  \begin{subfigure}{0.4\textwidth}
    \centering
    \begin{tikzpicture}[scale=1.2]
      \node (0) [nodraw, label=below:{\tettextsize 2}] at (0,0) {};
\node (1) [nodraw, label=below:{\tettextsize 3}] at (2.4,0) {};
\node (2) [nodraw, label=above right:{\tettextsize 4}] at (2.3,0.84) {};
\node (3) [nodraw, label=above:{\tettextsize 1}] at (1.2,2.0) {};

\path[line] (0) edge node[label=below:{\tettextsize 9}]{} (1);
\path[line, dashed] (0) edge node[label=above:{\tettextsize 14}]{} (2);
\path[line] (1) edge node[label=right:{\tettextsize 12}]{} (2);
\draw[line] (0) edge node[label=above left:{\tettextsize 11}]{} (3);
\draw[line] (1) edge node[label=left:{\tettextsize 10}]{} (3);
\draw[line] (2) edge node[label=above right:{\tettextsize 13}]{} (3);
    \end{tikzpicture}
    \label{fig:tet_simple}
    \caption{Vertex and edge numbering}
  \end{subfigure}\begin{subfigure}{0.4\textwidth}
    \centering
    \begin{tikzpicture}[scale=1.2]
      \def\y{.79}
\def\x{.32}
\node (0) [cell] at (0,0) {0};
\node (1) [facet] at (-3*\x, \y) {5};
\node (2) [facet] at (-1*\x, \y) {6};
\node (3) [facet] at (1*\x, \y) {7};
\node (4) [facet] at (3*\x, \y) {8};
\node (5) [edge] at (-4*\x, 2*\y) {9};
\node (6) [edge] at (-2.4*\x, 2*\y) {10};
\node (7) [edge] at (-.8*\x, 2*\y) {11};
\node (8) [edge] at (.8*\x, 2*\y) {12};
\node (9) [edge] at (2.4*\x, 2*\y) {13};
\node (10) [edge] at (4*\x, 2*\y) {14};
\node (11) [vertex] at (-3*\x, 3*\y) {1};
\node (12) [vertex] at (-1*\x, 3*\y) {2};
\node (13) [vertex] at (1*\x, 3*\y) {3};
\node (14) [vertex] at (3*\x, 3*\y) {4};

\draw[line] (0) -- (1);
\draw[line] (0) -- (2);
\draw[line] (0) -- (3);
\draw[line] (0) -- (4);
\draw[line] (1) -- (5);
\draw[line] (1) -- (6);
\draw[line] (1) -- (7);
\draw[line] (2) -- (6);
\draw[line] (2) -- (8);
\draw[line] (2) -- (9);
\draw[line] (3) -- (7);
\draw[line] (3) -- (9);
\draw[line] (3) -- (10);
\draw[line] (4.north west) -- (5.south east);
\draw[line] (4) -- (8);
\draw[line] (4) -- (10);
\draw[line] (5) -- (12);
\draw[line] (5) -- (13);
\draw[line] (6) -- (11);
\draw[line] (6) -- (13);
\draw[line] (7) -- (11);
\draw[line] (7) -- (12);
\draw[line] (8) -- (13);
\draw[line] (8) -- (14);
\draw[line] (9) -- (11);
\draw[line] (9) -- (14);
\draw[line] (10) -- (12);
\draw[line] (10) -- (14);
    \end{tikzpicture}
    \caption{Connectivity of entities in a DAG}
    \label{fig:tet_numbering}
  \end{subfigure}

  \begin{subfigure}{0.4\textwidth}
    \centering
    \begin{tikzpicture}[scale=1.2]
      \def\y{.79}
\def\x{.32}
\node (0) [cell] at (0,0) {0};
\node (1) [facet] at (-3*\x, \y) {5};
\node (2) [facet] at (-1*\x, \y) {6};
\node (3) [facet] at (1*\x, \y) {7};
\node (4) [facet] at (3*\x, \y) {8};
\node (5) [edge] at (-4*\x, 2*\y) {9};
\node (6) [edge] at (-2.4*\x, 2*\y) {10};
\node (7) [edge] at (-.8*\x, 2*\y) {11};
\node (8) [edge] at (.8*\x, 2*\y) {12};
\node (9) [edge] at (2.4*\x, 2*\y) {13};
\node (10) [edge] at (4*\x, 2*\y) {14};
\node (11) [vertex] at (-3*\x, 3*\y) {1};
\node (12) [vertex] at (-1*\x, 3*\y) {2};
\node (13) [vertex] at (1*\x, 3*\y) {3};
\node (14) [vertex] at (3*\x, 3*\y) {4};

\draw[line] (0) -- (1);
\draw[line] (0) -- (2);
\draw[line] (0) -- (3);
\draw[line] (0) -- (4);
\draw[line] (1) -- (5);
\draw[line] (1) -- (6);
\draw[line] (1) -- (7);
\draw[line] (2) -- (6);
\draw[line] (2) -- (8);
\draw[line] (2) -- (9);
\draw[line] (3) -- (7);
\draw[line] (3) -- (9);
\draw[line] (3) -- (10);
\draw[line] (4.north west) -- (5.south east);
\draw[line] (4) -- (8);
\draw[line] (4) -- (10);
\draw[line] (5) -- (12);
\draw[line] (5) -- (13);
\draw[line] (6) -- (11);
\draw[line] (6) -- (13);
\draw[line] (7) -- (11);
\draw[line] (7) -- (12);
\draw[line] (8) -- (13);
\draw[line] (8) -- (14);
\draw[line] (9) -- (11);
\draw[line] (9) -- (14);
\draw[line] (10) -- (12);
\draw[line] (10) -- (14);
      \begin{scope}[overlay, remember picture, rounded corners]
        \filldraw[fill opacity=0.2]
        ([xshift= 3pt, yshift=-3pt] 7.south east) --
        ([xshift= 3pt, yshift= 3pt] 7.north east) --
        ([xshift=-3pt, yshift= 3pt] 5.north west) --
        ([xshift=-3pt, yshift=-3pt] 5.south west) --
        cycle;
        \node[cell, fill=black, opacity=0.6] at (1.center) {};
      \end{scope}
    \end{tikzpicture}
    \caption{\centering $cone(5) = \linebreak\{9, 10, 11\}$}
    \label{fig:tet_cone}
  \end{subfigure}
  \begin{subfigure}{0.4\textwidth}\centering
    \begin{tikzpicture}[scale=1.2]
      \def\y{.79}
\def\x{.32}
\node (0) [cell] at (0,0) {0};
\node (1) [facet] at (-3*\x, \y) {5};
\node (2) [facet] at (-1*\x, \y) {6};
\node (3) [facet] at (1*\x, \y) {7};
\node (4) [facet] at (3*\x, \y) {8};
\node (5) [edge] at (-4*\x, 2*\y) {9};
\node (6) [edge] at (-2.4*\x, 2*\y) {10};
\node (7) [edge] at (-.8*\x, 2*\y) {11};
\node (8) [edge] at (.8*\x, 2*\y) {12};
\node (9) [edge] at (2.4*\x, 2*\y) {13};
\node (10) [edge] at (4*\x, 2*\y) {14};
\node (11) [vertex] at (-3*\x, 3*\y) {1};
\node (12) [vertex] at (-1*\x, 3*\y) {2};
\node (13) [vertex] at (1*\x, 3*\y) {3};
\node (14) [vertex] at (3*\x, 3*\y) {4};

\draw[line] (0) -- (1);
\draw[line] (0) -- (2);
\draw[line] (0) -- (3);
\draw[line] (0) -- (4);
\draw[line] (1) -- (5);
\draw[line] (1) -- (6);
\draw[line] (1) -- (7);
\draw[line] (2) -- (6);
\draw[line] (2) -- (8);
\draw[line] (2) -- (9);
\draw[line] (3) -- (7);
\draw[line] (3) -- (9);
\draw[line] (3) -- (10);
\draw[line] (4.north west) -- (5.south east);
\draw[line] (4) -- (8);
\draw[line] (4) -- (10);
\draw[line] (5) -- (12);
\draw[line] (5) -- (13);
\draw[line] (6) -- (11);
\draw[line] (6) -- (13);
\draw[line] (7) -- (11);
\draw[line] (7) -- (12);
\draw[line] (8) -- (13);
\draw[line] (8) -- (14);
\draw[line] (9) -- (11);
\draw[line] (9) -- (14);
\draw[line] (10) -- (12);
\draw[line] (10) -- (14);
      \begin{scope}[overlay, remember picture, rounded corners]
        \filldraw[fill opacity=0.2]
        ([xshift= 1pt, yshift=-3pt] 1.south east) --
        ([xshift=-3pt, yshift= 1pt] 2.north west) --
        ([xshift= 3pt, yshift=-1pt] 7.south east) --
        ([xshift=-3pt, yshift= 1pt] 8.north west) --
        ([xshift= 3pt, yshift=-1pt] 13.south east) --
        ([xshift= 3pt, yshift= 3pt] 13.north east) --
        ([xshift=-2pt, yshift= 3pt] 11.north west) --
        ([xshift=-2pt, yshift= 0pt] 5.west) --
        ([xshift=-1pt, yshift=-3pt] 1.south west) --
        cycle;
        \node[cell, fill=black, opacity=0.6] at (1.center) {};
      \end{scope}
    \end{tikzpicture}
    \caption{\centering $closure(5) = \linebreak\{1, 2, 3, 9, 10, 11\}$}
    \label{fig:tet_closure}
  \end{subfigure}

  \begin{subfigure}{0.4\textwidth}\centering
    \begin{tikzpicture}[scale=1.2]
      \def\y{.79}
\def\x{.32}
\node (0) [cell] at (0,0) {0};
\node (1) [facet] at (-3*\x, \y) {5};
\node (2) [facet] at (-1*\x, \y) {6};
\node (3) [facet] at (1*\x, \y) {7};
\node (4) [facet] at (3*\x, \y) {8};
\node (5) [edge] at (-4*\x, 2*\y) {9};
\node (6) [edge] at (-2.4*\x, 2*\y) {10};
\node (7) [edge] at (-.8*\x, 2*\y) {11};
\node (8) [edge] at (.8*\x, 2*\y) {12};
\node (9) [edge] at (2.4*\x, 2*\y) {13};
\node (10) [edge] at (4*\x, 2*\y) {14};
\node (11) [vertex] at (-3*\x, 3*\y) {1};
\node (12) [vertex] at (-1*\x, 3*\y) {2};
\node (13) [vertex] at (1*\x, 3*\y) {3};
\node (14) [vertex] at (3*\x, 3*\y) {4};

\draw[line] (0) -- (1);
\draw[line] (0) -- (2);
\draw[line] (0) -- (3);
\draw[line] (0) -- (4);
\draw[line] (1) -- (5);
\draw[line] (1) -- (6);
\draw[line] (1) -- (7);
\draw[line] (2) -- (6);
\draw[line] (2) -- (8);
\draw[line] (2) -- (9);
\draw[line] (3) -- (7);
\draw[line] (3) -- (9);
\draw[line] (3) -- (10);
\draw[line] (4.north west) -- (5.south east);
\draw[line] (4) -- (8);
\draw[line] (4) -- (10);
\draw[line] (5) -- (12);
\draw[line] (5) -- (13);
\draw[line] (6) -- (11);
\draw[line] (6) -- (13);
\draw[line] (7) -- (11);
\draw[line] (7) -- (12);
\draw[line] (8) -- (13);
\draw[line] (8) -- (14);
\draw[line] (9) -- (11);
\draw[line] (9) -- (14);
\draw[line] (10) -- (12);
\draw[line] (10) -- (14);
      \begin{scope}[overlay, remember picture, rounded corners]
        \filldraw[fill opacity=0.2]
        ([xshift= 3pt, yshift=-3pt] 10.south east) --
        ([xshift= 3pt, yshift= 3pt] 10.north east) --
        ([xshift=-3pt, yshift= 3pt] 8.north west) --
        ([xshift=-3pt, yshift=-3pt] 8.south west) --
        cycle;
        \node[cell, fill=black, opacity=0.6] at (14.center) {};
      \end{scope}
    \end{tikzpicture}
    \caption{\centering $support(4) = \linebreak\{12, 13, 14\}$}
    \label{fig:tet_support}
  \end{subfigure}
  \begin{subfigure}{0.4\textwidth}\centering
    \begin{tikzpicture}[scale=1.2]
      \def\y{.79}
\def\x{.32}
\node (0) [cell] at (0,0) {0};
\node (1) [facet] at (-3*\x, \y) {5};
\node (2) [facet] at (-1*\x, \y) {6};
\node (3) [facet] at (1*\x, \y) {7};
\node (4) [facet] at (3*\x, \y) {8};
\node (5) [edge] at (-4*\x, 2*\y) {9};
\node (6) [edge] at (-2.4*\x, 2*\y) {10};
\node (7) [edge] at (-.8*\x, 2*\y) {11};
\node (8) [edge] at (.8*\x, 2*\y) {12};
\node (9) [edge] at (2.4*\x, 2*\y) {13};
\node (10) [edge] at (4*\x, 2*\y) {14};
\node (11) [vertex] at (-3*\x, 3*\y) {1};
\node (12) [vertex] at (-1*\x, 3*\y) {2};
\node (13) [vertex] at (1*\x, 3*\y) {3};
\node (14) [vertex] at (3*\x, 3*\y) {4};

\draw[line] (0) -- (1);
\draw[line] (0) -- (2);
\draw[line] (0) -- (3);
\draw[line] (0) -- (4);
\draw[line] (1) -- (5);
\draw[line] (1) -- (6);
\draw[line] (1) -- (7);
\draw[line] (2) -- (6);
\draw[line] (2) -- (8);
\draw[line] (2) -- (9);
\draw[line] (3) -- (7);
\draw[line] (3) -- (9);
\draw[line] (3) -- (10);
\draw[line] (4.north west) -- (5.south east);
\draw[line] (4) -- (8);
\draw[line] (4) -- (10);
\draw[line] (5) -- (12);
\draw[line] (5) -- (13);
\draw[line] (6) -- (11);
\draw[line] (6) -- (13);
\draw[line] (7) -- (11);
\draw[line] (7) -- (12);
\draw[line] (8) -- (13);
\draw[line] (8) -- (14);
\draw[line] (9) -- (11);
\draw[line] (9) -- (14);
\draw[line] (10) -- (12);
\draw[line] (10) -- (14);
      \begin{scope}[overlay, remember picture, rounded corners]
        \filldraw[fill opacity=0.2]
        ([xshift= 0pt, yshift= 3pt] 14.north west) --
        ([xshift= 3pt, yshift=-3pt] 13.south east) --
        ([xshift=-3pt, yshift= 1pt] 8.north west) --
        ([xshift= 3pt, yshift=-1pt] 7.south east) --
        ([xshift=-3pt, yshift= 1pt] 2.north west) --
        ([xshift=-3pt, yshift= 1pt] 2.south west) --
        ([xshift= 0pt, yshift=-3pt] 0.south west) --
        ([xshift= 0pt, yshift=-3pt] 0.south east) --
        ([xshift= 3pt, yshift=-1pt] 4.south east) --
        ([xshift= 3pt, yshift= 0pt] 10.east) --
        ([xshift= 2pt, yshift= 3pt] 14.north east) --
        cycle;
        \node[cell, fill=black, opacity=0.6] at (14.center) {};
      \end{scope}
    \end{tikzpicture}
    \caption{\centering $star(4) = \linebreak\{0, 6, 7, 8, 12, 13, 14\}$}
    \label{fig:tet_star}
  \end{subfigure}

  \caption{Example entity numbering for a single tetrahedron and the
    corresponding internal DAG. Entities are numbered accross
    stratified layers (dimensions) with a consecutive numbering in
    each stratum.}
  \label{fig:plex_overview}
\end{figure}
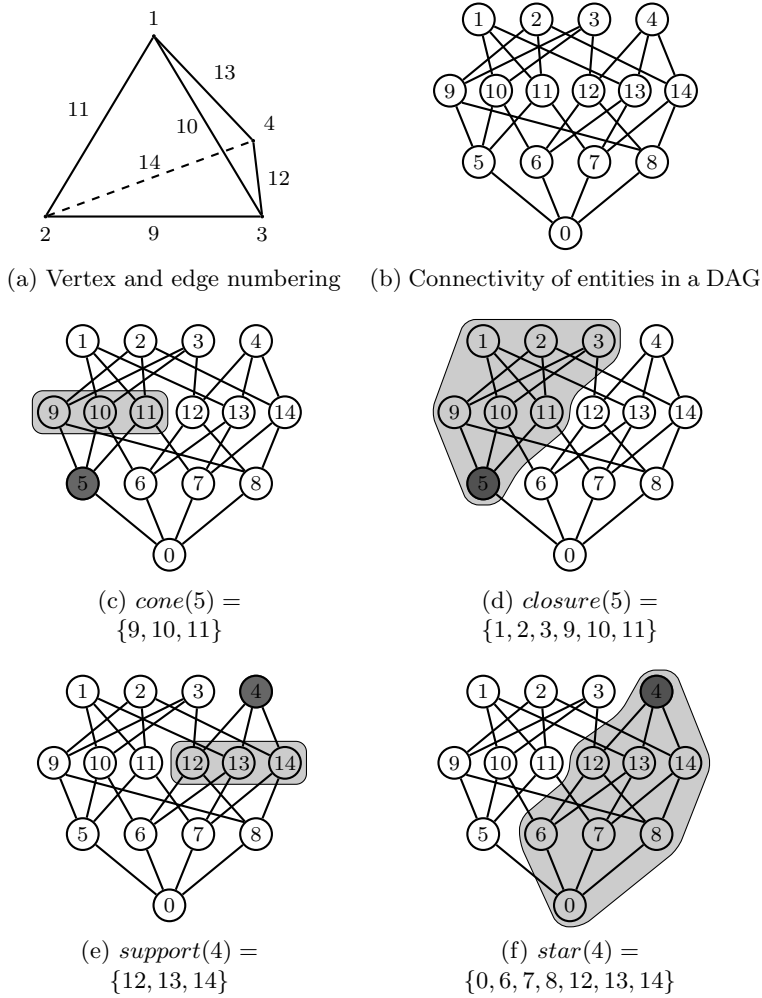

DMPlex uses an abstract representation of the unstructured meshes in
memory, where the connectivity of topological entities is stored as a
directed acyclic graph (DAG)~\cite{Knepley2009,Logg2009}. The DAG is
constructed of clearly defined layers (strata) that enable access to
mesh entities by their topological dimension or co-dimension, enabling
application codes to be written without explicit reference to the
topological dimension of the mesh. As illustrated in
\figref{fig:tet_numbering}, all points in the topology DAG share a
single consecutive entity numbering, emphasizing that each point is
treated equally no matter its shape or dimension, and allowing DMPlex to store the
graph connectivity in a single array where dimensional layers are
defined as consecutively numbered sub-ranges. The directional
connectivity of the DAG is defined by the covering relationship
$cone(p)$, which denotes all points directly connected to $p$ in the
next codimension, as illustrated in \figref{fig:tet_cone}. The
transitive closure of the $cone$ operation is denoted as $closure(p)$,
and depicted in \figref{fig:tet_closure}. The dual operation,
$support(p)$, and it's transitive closure $star(p)$ are shown in
\figref{fig:tet_support} and \figref{fig:tet_star} respectively.

In addition to the abstract topology data, PETSc provides two utility
objects to describe the parallel data layout: a {\em Section} object
maps the graph-based topology information to discretised solution data
through an offset mapping very similar to the Compressed Sparse Row (CSR)
storage scheme, and the {\em Star Forest}~\cite{StarForest11} (SF) object
holds a one-sided description
of shared data in parallel. These data layout mappings allow DMPlex to
manage distributed solution data by automating the preallocation of
distributed vector and matrix data structures and performing halo
data exchanges. Moreover, by storing grid topology alongside
discretised solution data, DMPlex is able provide the mappings required
for sophisticated preconditioning algorithms, such as geometric
multigrid methods~\cite{Brune2013} and multi-block, or ``Fieldsplit'',
preconditioning for multi-physics problems~\cite{Brown2012}.

\subsection{Mesh Reordering Techniques}
\label{sec:rcm_reordering}

The run-time performance of geometry-based processing algorithms can
be significantly affected by the data layout of unstructured meshes
and sparse matrices due to caching effects. A number of mesh ordering
techniques exist that aim to increase the cache coherency of local
data, either through cache-aware or cache-oblivious
reordering~\cite{Yoon2005,Gunther2006,Haase2007}. Cache-oblivious
techniques aim to reduce the bandwidth of the resulting sparse matrix
and thus lower the number of cache misses incurred when traversing
local data regardless of the underlying caching architecture.

The Reverse Cuthill-McKee (RCM)
algorithm~\cite{Cuthill1969,George1981} represents a classic example
of a cache-oblivious mesh reordering. RCM is based on a variant of a
simplex breadth-first search of the mesh connectivity graph and yields
a fixed-size $n$ tuple that represents the new ordering
permutation. Alternative methods, such as space filling curve numberings,
may be used to create similar permutations from a given mesh topology
graph in order to further increase cache coherency.

\section{Computational Meshes in Firedrake}
\label{sec:meshes}

The Firedrake system comprises a stack of specialised components that
implement a set of multi-layered abstractions to provide automated
finite element computation from a high-level
specification~\cite{Rathgeber2015}. The role of the top-level
Firedrake layer is to marshall data between the various sub-components
and to provide the computation layers, PyOP2 and PETSc, with the maps
and data objects required to assemble and solve linear and non-linear
systems. The computational mesh is encapsulated in a
\lstinline+Mesh+ object that can either be read from file or generated in
memory for common geometry classes, such as squares, cubes or spheres.

A characteristic feature of the Firedrake execution stack is that
multiple discretisations of the same computational domain, represented
by the \lstinline+FunctionSpace+ class, may be derived dynamically at any
point during execution, which requires the topological connectivity of
the mesh to be stored in a separate object. Separating mesh topology
from the discretisation of the problem not only enables Firedrake to
exploit caching and data re-use with minimal replication at multiple
levels in the tool chain but also allows data layout optimisations to
be inherited for all derived discretisations without re-computation of
the mesh reordering scheme.

As shown in \figref{fig:mesh_abstractions} the Firedrake classes
\lstinline+Mesh+ and \lstinline+FunctionSpace+, which encapsulate mesh topology
and problem discretisation respectively, map naturally onto the
abstractions provided by PETSc's data management API. The \lstinline+Mesh+
class encapsulates the topological connectivity of the grid by storing
a DMPlex object alongside a Firedrake-specific application ordering, while
discretisation data given by the \lstinline+FunctionSpace+ class defines
the layout of local data stored in the \lstinline+Function+ object.

\tikzset{
  >=stealth',
  box/.style={
    draw, rectangle, rounded corners
  },
  data/.style={
    draw, rectangle, minimum height=12pt, minimum width=12pt
  },
  split/.style={
    rectangle split, rectangle split parts=2
  },
  bheader/.style={
    above, yshift=12pt, font=\sffamily\bfseries\Large
  },
  btype/.style={
    below right, xshift=4pt, yshift=-4pt, font=\sffamily\bfseries
  },
}

\ifmonochrome
\tikzstyle{firedrake} = [box, minimum height=14em, fill=black!20]
\tikzstyle{petsc} = [box, minimum height=11em, fill=black!35, outer sep=5pt]
\tikzstyle{fiat} = [petsc, fill=black!10]
\tikzstyle{op2} = [petsc, fill=black!10]
\tikzstyle{local} = [data, fill=black!70]
\tikzstyle{global} = [data, fill=black!50]
\else
\tikzstyle{firedrake} = [box, minimum height=14em, fill=orange!60]
\tikzstyle{petsc} = [box, minimum height=11em, fill=blue!60, outer sep=5pt]
\tikzstyle{fiat} = [petsc, fill=red!35]
\tikzstyle{op2} = [petsc, fill=green!90]
\tikzstyle{local} = [data, fill=red]
\tikzstyle{global} = [data, fill=red!40]
\fi

\begin{figure}[ht]\centering
  \begin{tikzpicture}[font=\sf\small]
    \node (m) [firedrake, minimum width=0.32\textwidth] {};
    \node [btype] at (m.north west) {Mesh};
    \node [bheader, yshift=1em] at (m.north) {Topology};

    \node (m_plex) [petsc, text width=0.2\textwidth,
      anchor=south west] at (m.south west) {};
    \node (m_plex_title) [btype] at (m_plex.north west) {DMPlex};
    \node (m_plex_content) [below right, text width=0.2\textwidth] at
    (m_plex_title.south west) {
      \flushleft\hrule\vspace{2mm}
      - File I/O\\
      - Partitioning\\
      - Distribution\\
      - Mesh Topology\\
      - Renumbering\\
    };

    \node (m_perm) [petsc, minimum width=0.05\textwidth,
      anchor=north west, yshift=-2em, minimum height=9em] at (m_plex.north east) {};
    \node (m_perm_title) [btype] at (m_perm.north west) {IS};
    \node (m_perm_content) [below right, rotate=-90, anchor=north west,
    font=\sffamily\bfseries\large] at (m_perm_title.south east) {Permutation};

    \path [dashed, thick] ([shift={(.3,1.4)}] m.north east) edge
    ([shift={(.3,0)}] m.south east);

    \node (fs) [firedrake, minimum width=0.3\textwidth, minimum height=9em,
      right=2.6em of m.north east, anchor=north west] {};
    \node [bheader, yshift=1em] at (fs.north) {Discretisation};
    \node (fs_title) [btype] at (fs.north west) {FunctionSpace};

    \node (fs_fiat) [fiat, minimum width=0.12\textwidth, minimum height=1.2em,
      above=0.0 of fs.north west, anchor=south west, font=\tt] {FIAT.Element};

    \node (fs_sec) [petsc, minimum width=0.26\textwidth,
      minimum height=2.6em, anchor=north west] at (fs_title.south west) {};
    \node (fs_sec_title) [btype] at (fs_sec.north west) {PetscSection};
    \node (fs_sec_content) [below right, text width=0.24\textwidth]
    at (fs_sec_title.south west) {DAG$\rightarrow$DoF};

    \node (op2map) [op2, minimum width=0.26\textwidth,
      minimum height=2.6em, anchor=north west] at (fs_sec.south west) {};
    \node (op2map_title) [btype] at (op2map.north west) {PyOP2.Map};
    \node (op2map_content) [below right, text width=0.24\textwidth]
    at (op2map_title.south west) {Cell$\rightarrow$DoF};

    \node (halo) [firedrake, minimum width=0.3\textwidth, minimum height=4.4em,
      below=.5em of fs.south west, anchor=north west] {};
    \node (halo_title) [btype] at (halo.north west) {Halo};

    \node (halo_sf) [petsc, minimum width=0.26\textwidth,
      minimum height=1.6em, anchor=north west] at (halo_title.south west) {};
    \node (halo_sf_title) [btype] at (halo_sf.north west) {PetscSF};

    \path [dashed, thick] ([shift={(.3,1.4)}] fs.north east) edge
    ([shift={(.3,0)}] halo.south east);

    \node (f) [firedrake, minimum width=0.16\textwidth,
      right=1.6em of fs.north east, anchor=north west] {};
    \node [bheader, yshift=1em] at (f.north) {Data};
    \node (f_title) [btype] at (f.north west) {Function};

    \node (f_vec) [petsc, minimum width=0.12\textwidth,
      anchor=north west] at (f_title.south west) {};
    \node (f_vec_title) [btype] at (f_vec.north west) {Vec};

    \node (v0) [local, below left, anchor=north west, xshift=-.6em] at
    (f_vec_title.south east) {};
    \node (v1) [local, below=0 of v0] {};
    \node (v2) [local, below=0 of v1] {};
    \node (v3) [local, below=0 of v2] {};
    \node (v4) [local, below=0 of v3] {};
    \node (v5) [global, below=0.2em of v4] {};
    \node (v6) [global, below=0 of v5] {};

    \node (f_local) [rotate=-90, anchor=north west,
      font=\sffamily, shift={(.6,.5)}] at (v0.north east) {Local};
    \node (f_local) [rotate=-90, anchor=north west,
      font=\sffamily, shift={(.26,.5)}] at (v4.north east) {Remote};

    \draw [thick,->] ([shift={(-.18,-0.48)}] m_plex.north east) --
    ([shift={(.18,-0.5)}] fs_sec.north west) coordinate[pos=.284]
    (cperm) coordinate[pos=.76] (chalo);
    \draw [thick, <-,>=diamond] ([shift={(0,-.18)}]m_perm.north) -- (cperm);
    \draw [thick,->] (chalo) -- ++(0,-3.24) -- ([shift={(.18,0)}] halo_sf.west);
    \draw [thick] ([shift={(.18,0)}] fs_fiat.west) -- ++(-.44,0) -- (chalo);
    \draw [thick,->] ([shift={(0,.18)}] fs_sec.south) -- ([shift={(0,-.18)}] op2map.north);
    \draw [thick,->] ([shift={(-.18,0)}] op2map.east) --([shift={(.18,.14)}] f_vec.west);
    \draw [thick] ([shift={(-.18,.18)}] halo_sf.east) -- ++(1.6,0) -- ++(0,1.47)
    coordinate[pos=0.064] (cv6) coordinate[pos=0.36] (cv5)
    coordinate[pos=0.704] (cv4) coordinate[pos=1.] (cv3);
    \draw [thick] (cv6) -- (v6.west);
    \draw [thick] (cv5) -- (v5.west);
    \draw [thick] (cv4) -- (v4.west);
    \draw [thick] (cv3) -- (v3.west);

  \end{tikzpicture}
  \caption{Mapping of data abstractions between Firedrake and
  PETSc: Firedrake's \lstinline+Mesh+ object encapsulates domain
  topology stored in a \lstinline+DMPlex+ object alongside an
  application numbering permutation. The choice
  of \lstinline+FunctionSpace+ defines the local data discretisation
  via a \lstinline+PetscSection+ that is used to generate the
  indirection maps required by PyOP2 for assembly computation. Halo
  communication is performed by a \lstinline+PetscSF+ object, which
  encapsulates the mapping between local and remote data items in the
  local \lstinline+Vec+.}
  \label{fig:mesh_abstractions}
\end{figure}
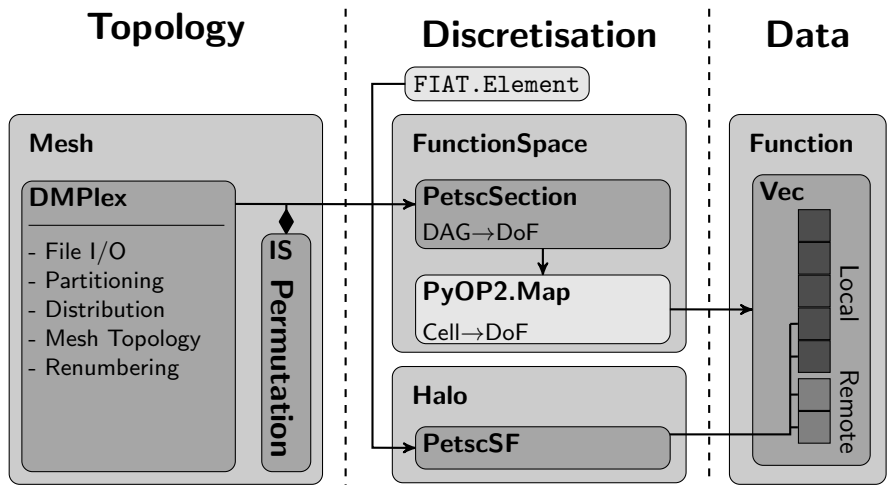

\subsection{Mesh topology}

Firedrake uses the DMPlex data management abstraction as an internal
representation of mesh topology, allowing it to delegate file I/O and
run-time mesh generation to PETSc. In doing so Firedrake only depends
on the public API provided by PETSc and automatically inherits the
mesh management and manipulation capabilities provided by DMPlex. As a
result Firedrake naturally supports the same set of mesh file formats
as DMPlex, which at the time of writing includes ExodusII, Gmsh, CGNS,
MED and Fluent Case files, and thus increases interoperability with
other applications and provides extensibility through a well-supported
public library.

In addition to various mesh format readers DMPlex also provides
parallel domain decomposition routines that interface with external
libraries, such as Chaco and Metis/ParMetis, to facilitate parallel
partitioning of the topology graph. Utilising PETSc's internal
communication routines DMPlex is thus capable of automatically
distributing the mesh across any number of processes, which allows
Firedrake to fully automate the parallelisation and optimisation of
the user-defined Finite Element problem.

Another advantage of using the DMPlex DAG as an intermediate
representation of mesh topology is that the abstracted graph format
allows Firedrake to dictate the ordering of the mesh topology and thus
control local data layout of derived discretisations. This is made
possible by attaching a point permutation to the DMPlex object, which
defines a single level of indirection that is applied to all graph
traversal operations within DMPlex. As a results, all discretisation
objects derived from the stored topology inherit this permutation,
giving Firedrake an effective way to control the global ordering of
derived solution data.

\subsection{Discretisation}

The FEniCS language (UFL~\cite{Alnaes2014}) implemented by Firedrake
allows the use of various discretisation schemes to represent solution
data, where the number of DoFs associated with each mesh entity is
determined by the local discretisation within a reference element. The
FIAT package~\cite{Kirby2004} of the FEniCS software stack provides
this reference element from which Firedrake needs to derive the
indirection {\em maps} between mesh cells and DoFs required by PyOP2
to perform matrix and vector assembly.

The mapping from mesh topology to solution data is facilitated by
PETSc through \lstinline+PetscSection+, a class of descriptor objects that
store a CSR-style mapping between points in the topology DAG and
entries in array or vector objects. Assuming a constant element type
throughout the mesh, DMPlex can generate a {\em section} object, given
the number of DoFs associated with each mesh entity type as provided
by the FIAT reference element. The set of DoFs associated with a cell
can then be derived by taking the {\em closure} of the cell point
(see \figref{fig:tet_closure}) and collecting the DoFs associated with
them by the provided {\em section}.

The use of DMPlex {\em closures} to determine entity-to-DoF mappings
is sufficient on its own should the local numbering of mesh entities
within a cell closure match that required by the application.
In Firedrake the local numbering on simplices must match the simplex
numbering used in FEniCS~\cite{Logg2012}, where the local
facet number is determined by the local number of the opposite
vertex. The algorithm shown in Alg.~\ref{alg:local_numbering} is thus
applied to each cell closure in turn to enforce the desired local
numbering for simplices.

\begin{algorithm}
  \begin{algorithmic}[1]
    \For{$cell$ in $mesh$}               \Comment{Loop over all cells in the mesh}
    \State $closure_{cell} \gets$ \Call{DMPlexGetClosure}{$plex, cell$}
    \For{$p$ in $closure_{cell}$}        \Comment{Filter facets and vertices from cell closure}
    \If{$p$ in \Call{DepthStratum}{$plex, 0$}} $vertices \gets p$\EndIf
    \If{$p$ in \Call{HeightStratum}{$plex, 1$}} $facets \gets p$\EndIf
    \EndFor
    \State \Call{Sort}{vertices}         \Comment{Sort vertices by global number}
    \For{$facet$ in $facets$}
    \State $closure_{facet} \gets$ \Call{DMPlexGetClosure}{$plex, facet$}
    \For{$f$ in $closure_{facet}$}       \Comment{Filter vertices from facet closure}
    \If{$f$ in \Call{DepthStratum}{$plex, 0$}} $v_{facet} \gets f$\EndIf
    \EndFor
    \For{$v$ in $vertices$}              \Comment{Find non-adjacent vertices}
    \If{$v$ not in $v_{facet}$} $keys \gets (facet, v)$\EndIf
    \EndFor
    \EndFor
    \State \Call{Sort}{$facets, keys$}   \Comment{Sort facets by non-adjacent vertices}
    \EndFor
  \end{algorithmic}
  \caption{Local numbering algorithm for simplex elements}
  \label{alg:local_numbering}
\end{algorithm}

\subsection{Halo communication}

The exchange of halo data between processors in Firedrake is performed
by PETSc's {\em Star Forest}~\cite{StarForest11} (SF) communication
abstraction that encapsulates one-sided description of shared data. SF
objects implement a range of sparse communication patterns that are
able to perform common data communication patterns, such broadcasts
and reduction operations, over sparse data arrays according to the
stored mapping. The halo data exchange pattern is derived by DMPlex
from an internal SF encapsulating the overlap in the topology graph
and a given discretisation provided in the form of a {\em section}
object. The derived SF encapsulates a local-to-local remote data
mapping that avoids the need to convert halo data into a global
numbering.

\section{Application orderings}

When deriving function spaces from a DMPlex topology definition, the
global data layout is inherited from the original graph ordering
generated by PETSc.  PyOP2, however, imposes a data layout restriction
that allows it to optimise performance by overlapping computation with
communication, which is not honoured in the global entity numbering
generated by DMPlex. Firedrake therefore generates an
\emph{application ordering} in the form of a permutation of the DAG
points that is passed to a distributed
DMPlex object to generate indirection maps that adhere to its required
ordering.

\subsection{PyOP2 data ordering}

To ensure that halo exchange communication can be overlapped
with assembly kernel computation PyOP2 {\em sets} require a strict
entity ordering, where non-owned data is stored contiguously at the
end of the data array. Moreover, as shown in
\figref{fig:mesh_ordering_pyop2}, all owned data items adjacent to
non-owned items require that the halo data exchange be finished before
computation is performed. Thus, owned data is further partitioned into
{\em core} (independent of halo) and {\em non-core} (halo-dependent)
data, allowing processing over {\em core} data items to proceed while
communication is still in-flight.

\tikzstyle{vertex} = [circle, inner sep=2pt]

\ifmonochrome
\colorlet{core}{black}
\colorlet{noncore}{black!70}
\colorlet{halo}{black!30}
\else
\colorlet{core}{orange}
\colorlet{noncore}{red}
\colorlet{halo}{blue}
\fi

\def\numEdges{4}

\newcommand{\drawVertex}[3] {
  \node[vertex, draw=black!20!#3, fill=black!20!#3] at (#1,#2) {};
}

\newcommand{\drawTriangleLower}[3] {
  \drawVertex {#1}{#2}{#3}
  \drawVertex {#1+1}{#2}{#3}
  \drawVertex {#1}{#2+1}{#3}
  \filldraw[thick, draw=black!20!#3, fill=black!20!#3, fill opacity=0.4]
  (#1,#2) -- (#1+1,#2) -- (#1,#2+1) --cycle;
}

\newcommand{\drawTriangleUpper}[3] {
  \drawVertex {#1}{#2}{#3}
  \drawVertex {#1-1}{#2}{#3}
  \drawVertex {#1}{#2-1}{#3}
  \filldraw[thick, draw=black!20!#3, fill=black!20!#3, fill opacity=0.4]
  (#1,#2) -- (#1-1,#2) -- (#1,#2-1) --cycle;
}

\newcommand{\drawGridLower}[3] {
  \foreach \x in {0, ..., #2} {
    \pgfmathsetmacro{\maxy}{#2 - \x}
    \pgfmathsetmacro{\miny}{max(0, #1 - \x}
    \foreach \y in {\miny, ..., \maxy} {
      \drawTriangleLower{\x}{\y}{#3}
    }
  }
  \pgfmathsetmacro{\maxx}{#2}
  \foreach \x in {1, ..., \maxx} {
    \pgfmathsetmacro{\maxy}{#2 - \x + 1}
    \pgfmathsetmacro{\miny}{max(1, #1 - \x + 1}
    \foreach \y in {\miny, ..., \maxy} {
      \drawTriangleUpper{\x}{\y}{#3}
    }
  }
}

\newcommand{\drawGridUpper}[3] {
  \foreach \x in {0, ..., #2} {
    \pgfmathsetmacro{\maxy}{#2 - \x}
    \pgfmathsetmacro{\miny}{max(0, #1 - \x}
    \foreach \y in {\miny, ..., \maxy} {
      \drawTriangleUpper{\numEdges-\x}{\numEdges-\y}{#3}
    }
  }
  \pgfmathsetmacro{\maxx}{#2}
  \foreach \x in {1, ..., \maxx} {
    \pgfmathsetmacro{\maxy}{#2 - \x + 1}
    \pgfmathsetmacro{\miny}{max(1, #1 - \x + 1}
    \foreach \y in {\miny, ..., \maxy} {
      \drawTriangleLower{\numEdges-\x}{\numEdges-\y}{#3}
    }
  }
}

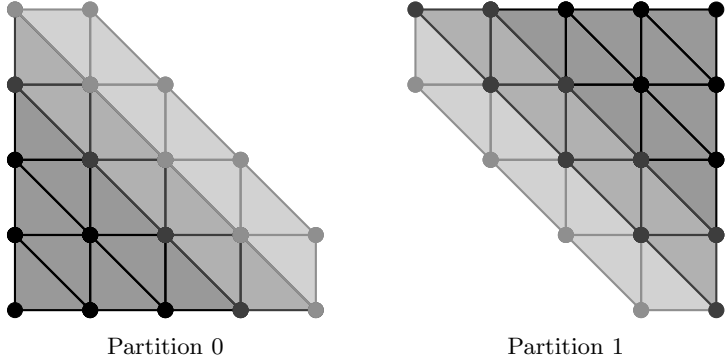
\begin{figure}[ht]\centering
  \begin{subfigure}{0.4\textwidth} \centering
    \begin{tikzpicture}
      \drawGridLower{0}{2}{core}
      \drawGridLower{3}{3}{noncore}
      \drawGridUpper{3}{3}{halo}

    \end{tikzpicture}
    \caption*{Partition 0}
  \end{subfigure}
  \begin{subfigure}{0.4\textwidth} \centering
    \begin{tikzpicture}
      \drawGridLower{3}{3}{halo}
      \drawGridUpper{0}{2}{core}
      \drawGridUpper{3}{3}{noncore}
    \end{tikzpicture}
    \caption*{Partition 1}
  \end{subfigure}

  \caption{PyOP2 entity classes on a distributed $4\times4$ unit
  square mesh. The dark region marks {\em core} entities, medium grey
  marks {\em non-core} entities and light grey marks the halo region.}
  \label{fig:mesh_ordering_pyop2}
\end{figure}

Firedrake honours the PyOP2 entity ordering by assigning all points in
the DMPlex topology DAG to one of the PyOP2 entity classes using a
\lstinline+DMLabel+ data structure, which encapsulates integer value
assignments to points. When deriving the indirection maps for each discretisation,
mesh entities can then be filtered into the appropriate sets
regardless of entity type. The algorithm used to mark PyOP2 entity
classes is shown in Alg.~\ref{alg:entity_marking}, where the initial
overlap definition, provided by DMPlex in form of an SF, is used to
first mark the {\em halo} region, followed by the derivation of
adjacent {\em non-core} points.

\begin{algorithm}
  \begin{algorithmic}[1]
    \For{$p$ in $pointSF$}                      \Comment{Define halo region from SF}
    \State \Call{LabelSetValue}{$halo, p$}
    \EndFor
    \For{$p$ in \Call{LabelGetStratum}{$halo$}} \Comment{Loop over halo cells}
    \If{$p$ in \Call{HeightStratum}{$plex, 0$}}
    \State $adjacency \gets$ \Call{DMPlexGetAdjacency}{$plex, p$}
    \For{$c$ in $adjacency$}                    \Comment{Find cells adjacent to halo}
    \If{\Call{LabelHasPoint}{$halo$, c} and $c$ in \Call{HeightStratum}{$plex, 0$}}
    \State \Call{LabelSetValue}{$noncore, p$}   \Comment{Mark adjacent cell as non-core}
    \EndIf\EndFor
    \EndIf\EndFor
    \For{$p$ in $mesh$}                         \Comment{Mark remaining points as core}
    \If{not \Call{LabelHasPoint}{$halo$, p} and not \Call{LabelHasPoint}{$noncore$, p}}
    \State \Call{LabelSetValue}{$core, p$}
    \EndIf\EndFor
  \end{algorithmic}
  \caption{Algorithm to mark PyOP2 entity classes on DMPlex based on
    the initial {\em halo} definition given by DMPlex. Point adjacency
    in the DAG is defined as $adjacency(p) = closure(star(p))$.}
  \label{alg:entity_marking}
\end{algorithm}

\subsection{Compact RCM ordering}

The generic encapsulation of mesh topology allows DMPlex to compute
the point permutation according to the well-known RCM mesh reordering
algorithm (see \secref{sec:rcm_reordering}). Since Firedrake already
controls the effective ordering of mesh entities to adhere to
PyOP2 ordering restrictions, the RCM permutation provided by DMPlex
can be applied to the Firedrake-specific point permutation. However,
any additional indirection applied to the reordering permutation
computed by Firedrake needs to be contained within the marked PyOP2
class regions. Thus, although the base RCM permutation generated by
DMPlex includes all graph points, Firedrake implements a cell-wise
compact reordering, where the cell ordering is filtered from the RCM
permutation within each marked PyOP2 region. As shown in
Alg.~\ref{alg:compact_rcm}, the full permutation is then derived by
adding cell {\em closures} along the segmented cell order, ensuring
the relative compactness of DoFs associated with the same cell.

\begin{algorithm}
  \begin{algorithmic}[1]
    \State{$ordering \gets$ \Call{DMPlexGetOrdering}{$RCM$}} \Comment{Get RCM renumbering}
    \For{$class$ in $\{core, noncore, halo\}$}           \Comment{Get array index for each class}
    \State $idx_{class} \gets$ \Call{LabelStratumSize}{$class$}
    \EndFor
    \For{$p$ in $mesh$}
    \State $p_{rcm} \gets ordering\{p\}$
    \If{$p_{rcm}$ not in \Call{HeightStratum}{$plex, 0$}} skip $p$ \EndIf
    \For{$class$ in $\{core, noncore, halo\}$}           \Comment{Get array index for current class}
    \If{\Call{LabelHasPoint}{$class, p$}} $idx \gets idx_{class}$ \EndIf
    \For{$p_{closure}$ in \Call{DMPlexGetClosure}{$plex, p_{rcm}$}}
      \State $permutation\{idx\} \gets p_{closure}$
    \EndFor
    \EndFor
    \EndFor
  \end{algorithmic}
  \caption{Algorithm for generating a compact RCM permutation that
    honours PyOP2's entity class separation and encapsulates a
    cell-wise RCM reordering within the PyOP2 regions.}
  \label{alg:compact_rcm}
\end{algorithm}

\begin{figure}[ht]
  \centering
  \begin{subfigure}{0.4\textwidth}\centering
    \fbox{\includegraphics[width=0.8\textwidth]
      {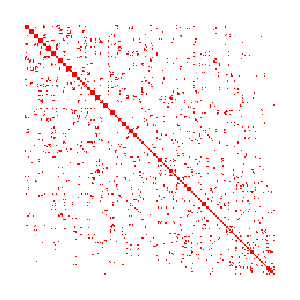}}
    \caption{$P_1$, native, sequential}
  \end{subfigure}\begin{subfigure}{0.4\textwidth}
    \centering
    \fbox{\includegraphics[width=0.8\textwidth]
      {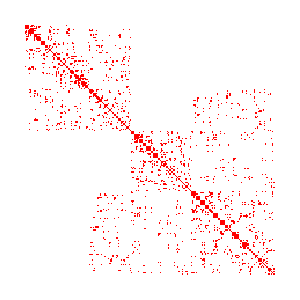}}
    \caption{$P1$, native, parallel}
  \end{subfigure}
  \vspace{4mm}

  \begin{subfigure}{0.4\textwidth}\centering
    \fbox{\includegraphics[width=0.8\textwidth]
      {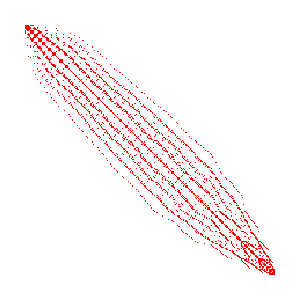}}
    \caption{$P_1$, RCM, sequential}
  \end{subfigure}\begin{subfigure}{0.4\textwidth}\centering
    \fbox{\includegraphics[width=0.8\textwidth]
      {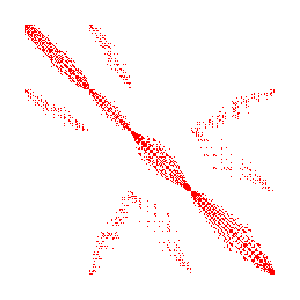}}
    \caption{$P_1$, RCM, parallel}
  \end{subfigure}
  \vspace{4mm}

  \begin{subfigure}{0.4\textwidth}\centering
    \fbox{\includegraphics[width=0.8\textwidth]
      {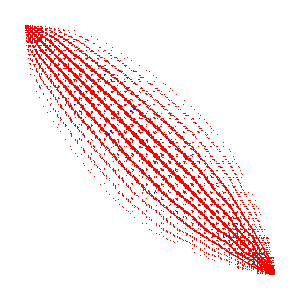}}
    \caption{$P_3$, RCM, sequential}
  \end{subfigure}\begin{subfigure}{0.4\textwidth}\centering
    \fbox{\includegraphics[width=0.8\textwidth]
      {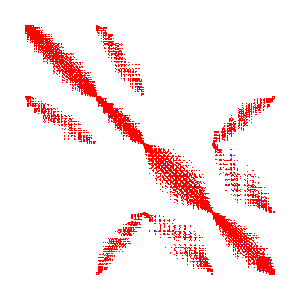}}
    \caption{$P_3$, RCM, parallel}
  \end{subfigure}

  \caption{Effects of the combined RCM and OP2 mesh ordering on matrix
    structure for a $P_1$ and a $P_3$ function space on a $5\times5$
    unit square.}
  \label{fig:matrix_reordering}
\end{figure}

It is worth noting that this cell-wise compact reordering approach
allows any additional level of indirection to be applied without
violating the PyOP2 ordering constraint and is therefore not limited
to RCM. Examples of sparse matrix structures generated using the
RCM-based reordering are shown in \figref{fig:matrix_reordering}.

\section{Performance benchmarks}

The benefits of Firedrake's compact RCM mesh reordering have been
evaluated using two sets of performance benchmarks: a run-time
comparison of assembly loops over cells and interior facets with
light-weight kernels, as well as solving a full advection-diffusion
problem. The benchmark experiments were carried out on the UK national
supercomputer ARCHER, a Cray XE30 with 4920 nodes connected via an
Aries interconnect and a parallel Lustre
filesystem~\footnote{\url{http://www.archer.ac.uk/}}. Each node
consists of two 2.7 GHz, 12-core Intel E5-2697 v2 (Ivy Bridge)
processors with 64GB of memory.

An indication of the indirection cost and subsequent data traversal
performance in low-level loops was gained by comparing the
individually measured execution time of two PyOP2 assembly loops. The
benchmark loops were generated by invoking \lstinline+assemble(L)+ 100
times for the UFL expressions \lstinline+L = u*dx+ and
\lstinline!L = u('+')*dS! for cell and interior facet integrals
respectively, where \lstinline+u+ is a suitable \lstinline+Function+
object. The performance of a full-scale finite element problem was
then analysed, which consisted of assembling and solving the
advection-diffusion equation $\frac{\partial c}{\partial
  t}+\bigtriangledown\cdotp(\vec{u}c) =
\bigtriangledown\cdotp(\overline{\overline{\kappa}}\bigtriangledown
c)$ using a Conjugate Gradient method with a Jacobi preconditioner for
advection and the HYPRE BoomerAMG algebraic multigrid
preconditioner~\cite{Falgout2006} for the diffusion component. The
mesh used in both experiments represents a two-dimensional L-shaped
domain, consisting of 3,105,620 cells and 1,552,808 vertices, and was
generated with Gmsh~\cite{Geuz2009}.

\begin{figure}[ht]
  \centering
  \begin{subfigure}{0.5\textwidth}\centering
    \includegraphics[width=0.95\textwidth, clip=true]
      {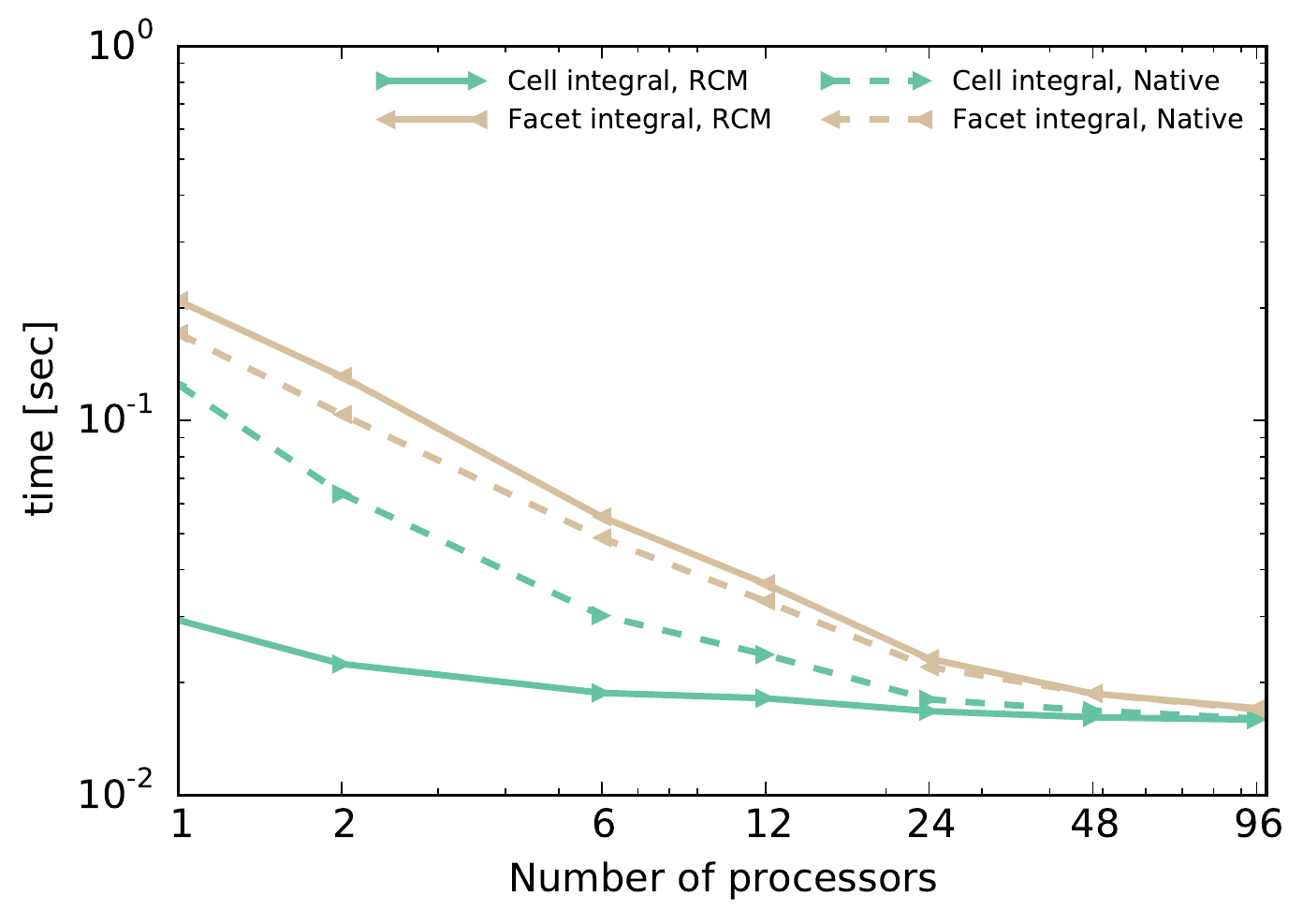}
    \caption{PyOP2 loop performance for $P_1$}
  \end{subfigure}\begin{subfigure}{0.5\textwidth}\centering
    \includegraphics[width=0.95\textwidth, clip=true]
      {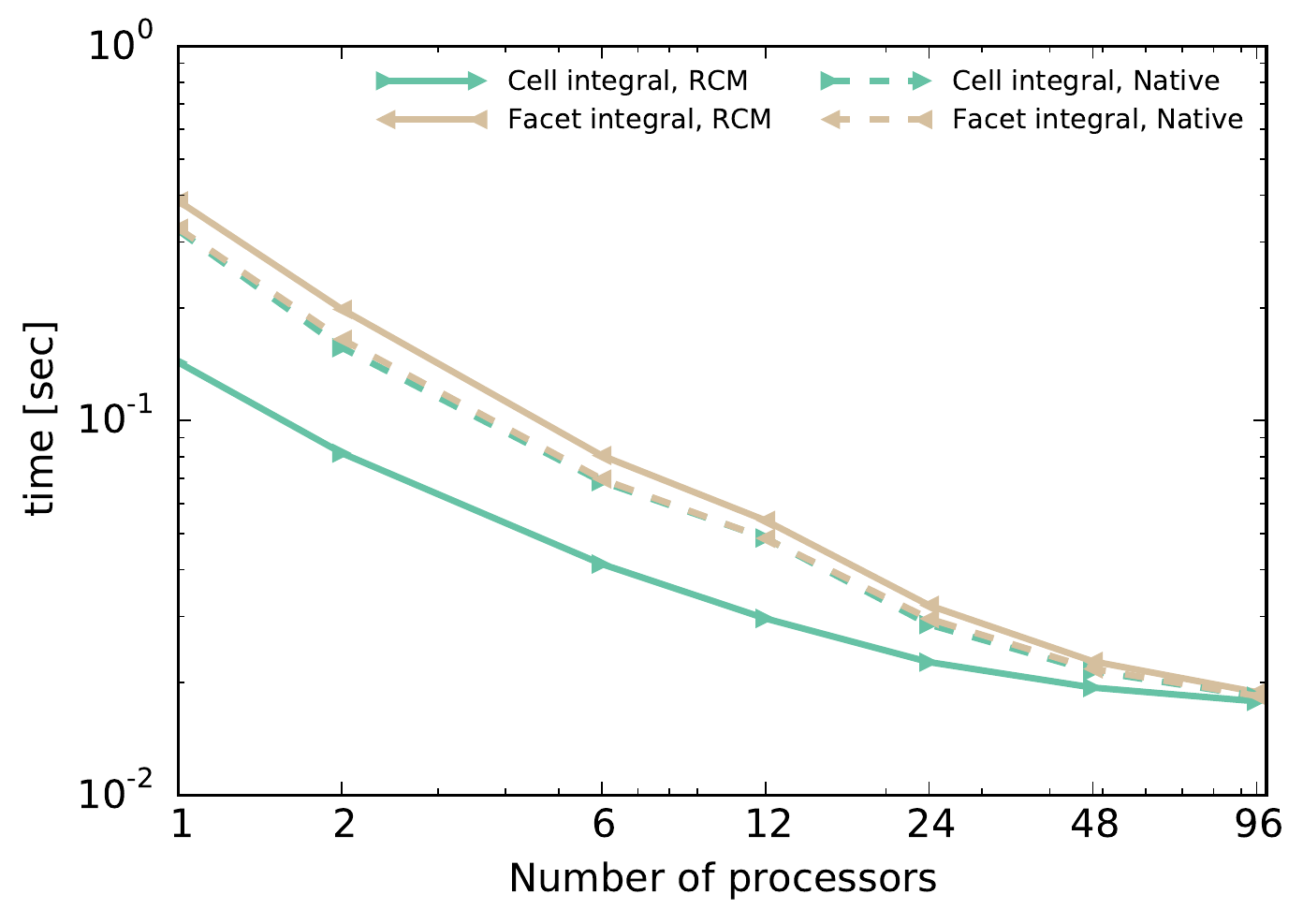}
    \caption{PyOP2 loop performance for $P_3$}
  \end{subfigure}
  \caption{Run-time comparison between compact RCM and native
    numbering for assembly loops over cells and interior
    facets.}
  \label{fig:results_ordering}
\end{figure}

The performance of the assembly loops over cells and interior
facets using $P_1$ and $P_3$ function spaces on up to 96 cores is
shown in \figref{fig:results_ordering}. The performance of the cell
integral loop shows significant improvements in both cases, whereas the
facet integral loop shows a small performance decrease. This
highlights that the compact RCM reordering optimises cell integral
computation due to the generated cell-wise compact traversal
pattern. It is also worth noting that the improvement due to RCM
diminishes as we approach the strong scaling limit, although an
increase in computational intensity between $P_1$ and $P_3$ assembly
kernels negates this effect.

\begin{figure}[ht]
  \centering
  \begin{subfigure}{0.5\textwidth}\centering
    \includegraphics[width=0.95\textwidth, clip=true]
      {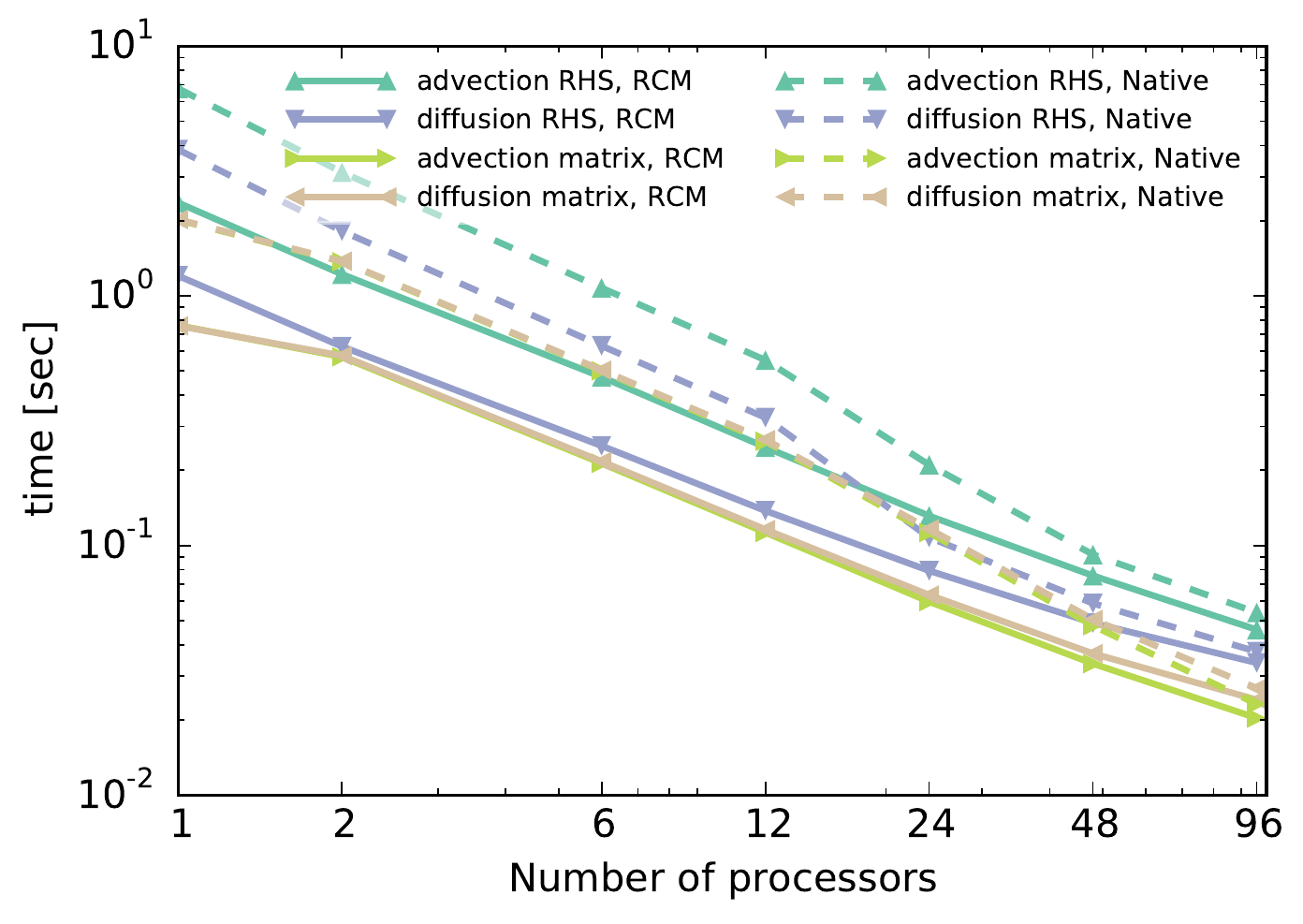}
    \caption{Assembly performance for $P_1$}
    \label{fig:results_adv_diff_assemble_p1}
  \end{subfigure}\begin{subfigure}{0.5\textwidth}\centering
    \includegraphics[width=0.95\textwidth, clip=true]
      {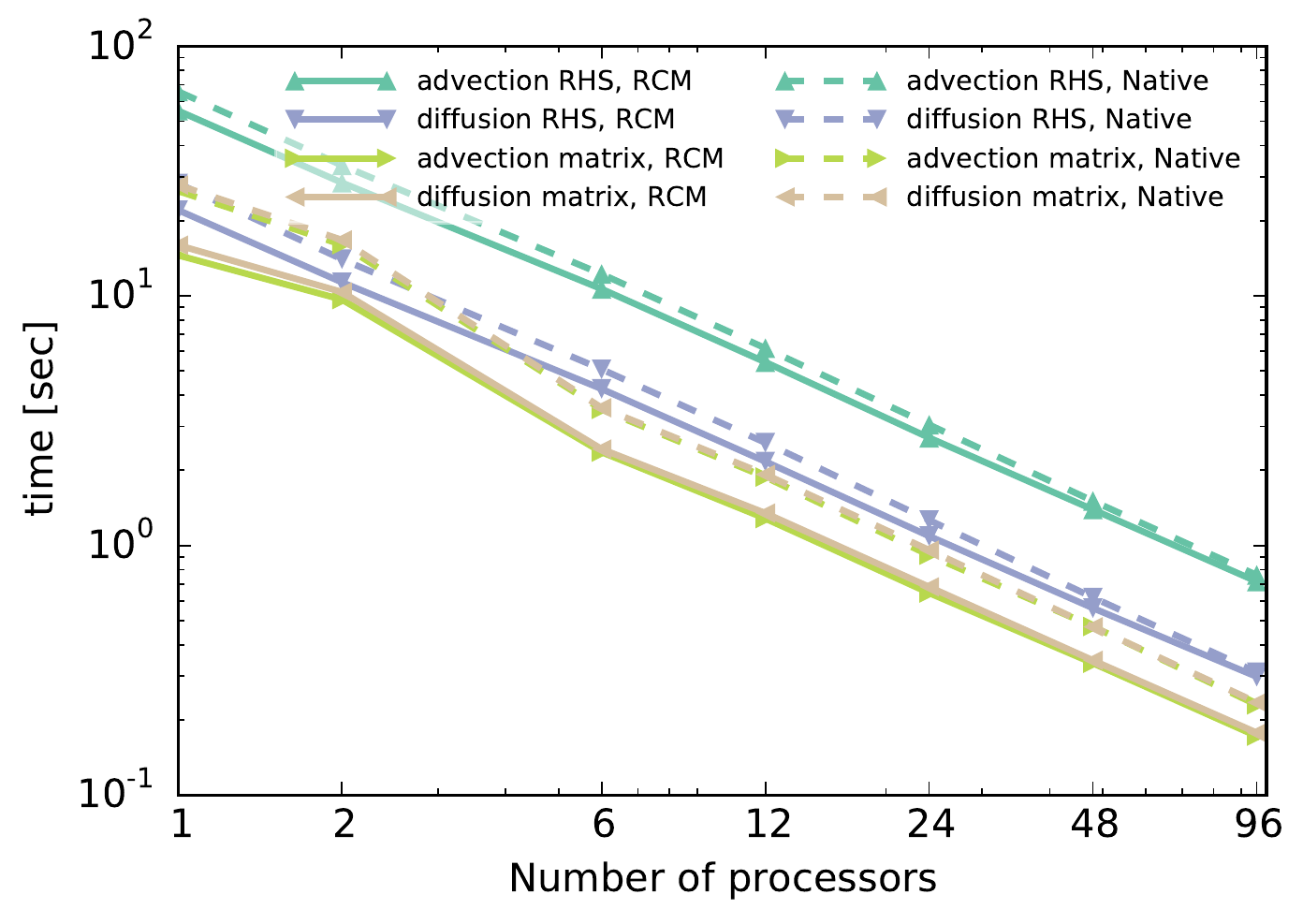}
    \caption{Assembly performance for $P_3$}
    \label{fig:results_adv_diff_assemble_p3}
  \end{subfigure}
  \vspace{2mm}

  \begin{subfigure}{0.5\textwidth}\centering
    \includegraphics[width=0.95\textwidth, clip=true]
      {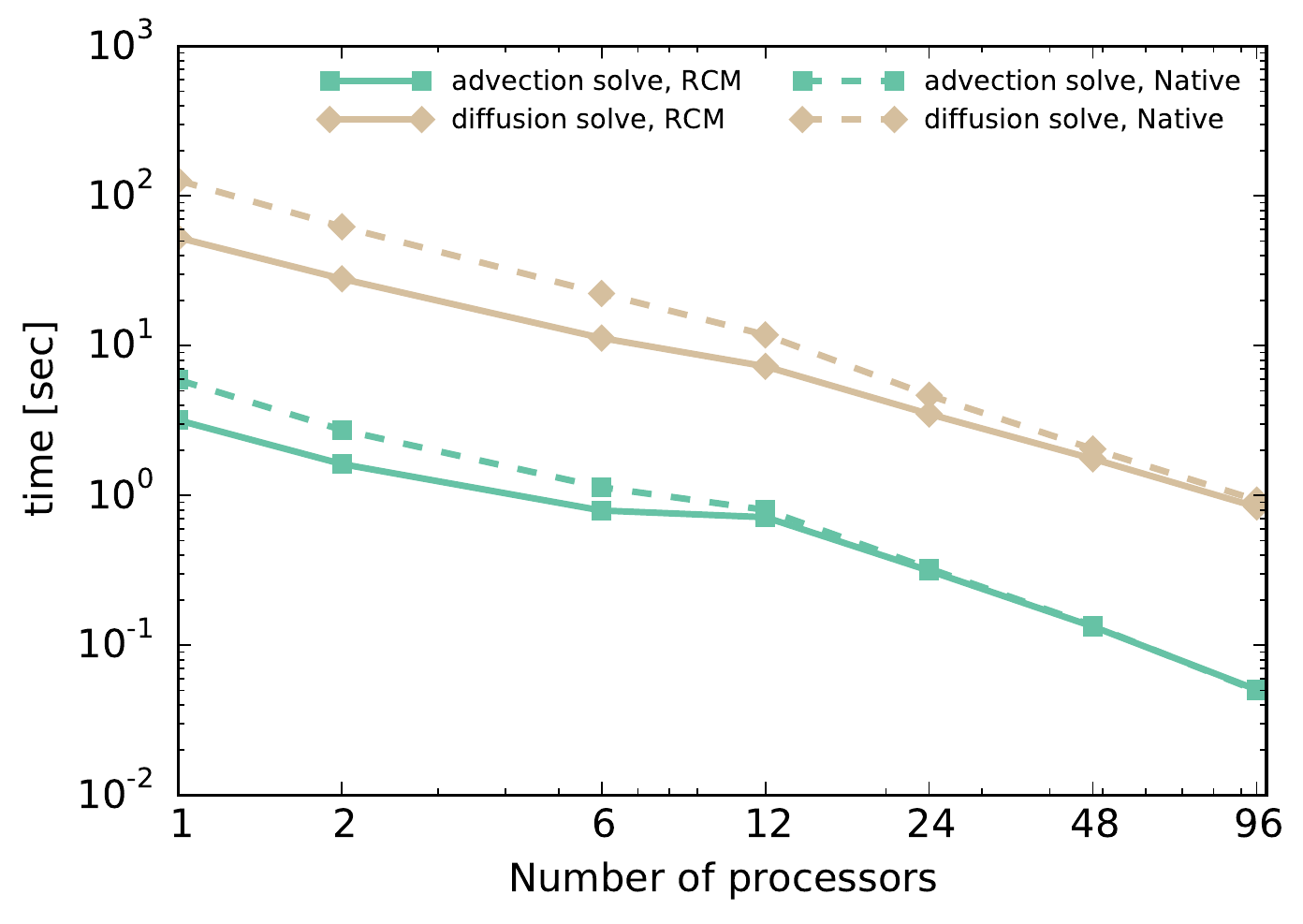}
    \caption{Solver performance for $P_1$}
    \label{fig:results_adv_diff_solve_p1}
  \end{subfigure}\begin{subfigure}{0.5\textwidth}\centering
    \includegraphics[width=0.95\textwidth, clip=true]
      {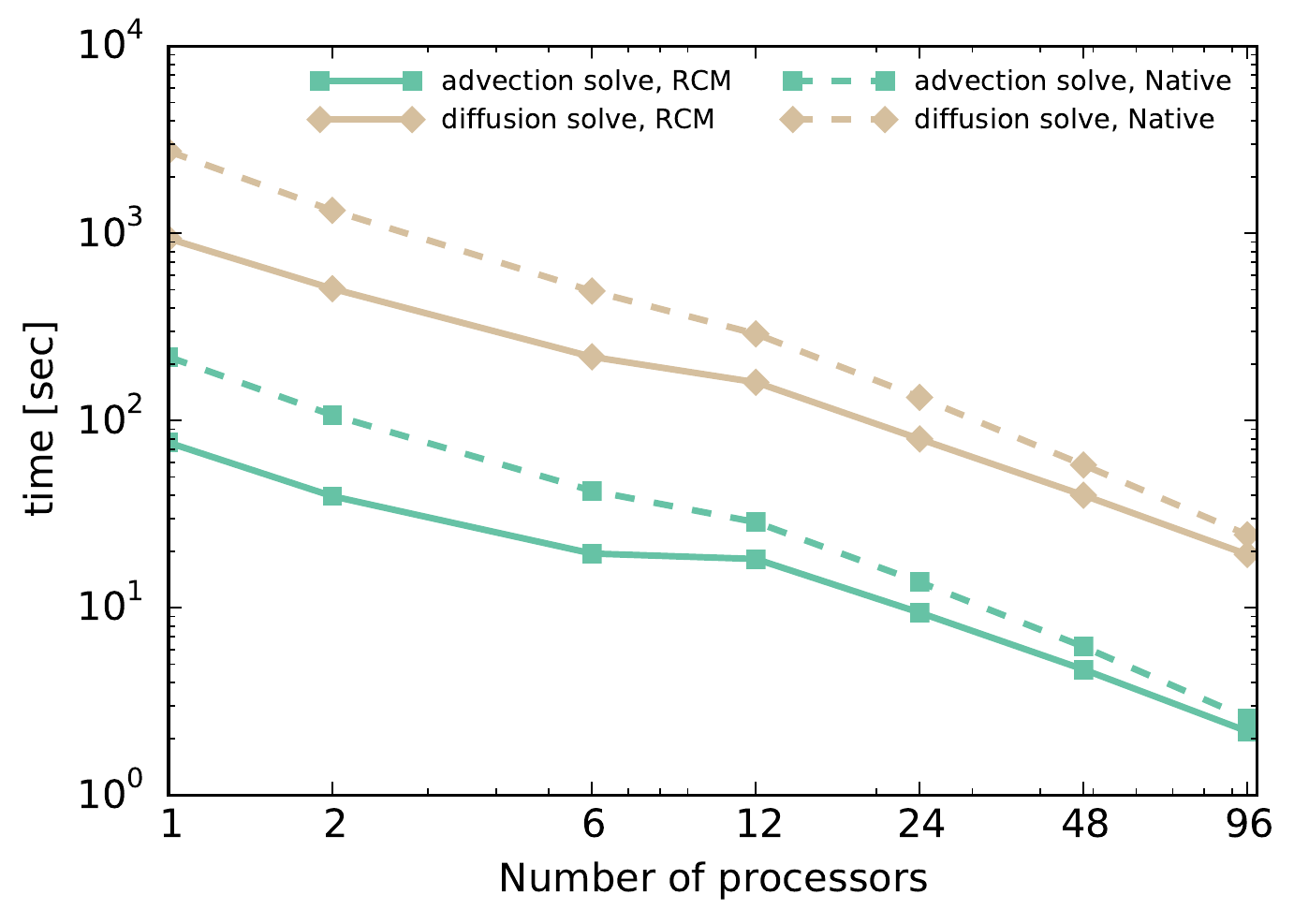}
    \caption{Solver performance for $P_3$}
    \label{fig:results_adv_diff_solve_p3}
  \end{subfigure}
  \caption{Run-time comparison between compact RCM and native
    numbering for the advection-diffusion problem on $P_1$ and $P_3$
    discretisations.}
  \label{fig:results_adv_diff}
\end{figure}

A performance profile of the full advection-diffusion model is given
in \figref{fig:results_adv_diff}. Matrix and RHS assembly times
indicate clear performance improvements under compact RCM with
significant speedups for $P_1$ on small numbers of cores (see
\figref{fig:results_adv_diff_assemble_p1}). As shown in
\figref{fig:results_adv_diff_assemble_p3}, $P_3$ assembly kernels with
a higher computational intensity also show significant performance
improvements, where matrix assembly in particular benefits from the
reordering in a sustained way up to 96 cores. Similarly, advection and
diffusion solver times shown in \figref{fig:results_adv_diff_solve_p1}
and \figref{fig:results_adv_diff_solve_p3} indicate a clear speedup on
small numbers of cores, while significant improvements are also
evident on up to 96 cores for solves with larger numbers of DoFs in
$P_3$.

\section{Discussion}

In this paper we give a full account of the utilisation of PETSc's
DMPlex topology abstraction in Firedrake to derive the topological
mapping required to solve a wide range of finite element problems. We
highlight how the right composition of abstractions can be used to
apply well known data layout optimisations, such as RCM renumbering,
to an entire class of problems and demonstrate the resulting gains in
assembly and solver performance. Our work emphasises the importance of
high-level DSLs and further underlines their potential for achieving
performance portability through run-time optimisation.

An important corollary of the close integration of DMPlex into the
Firedrake framework is the improved interoperability and extensibility
of the mesh management component. Future efforts to improve file I/O
and add new meshing capabilities, such as mesh adaptivity, can now be
integrated through PETSc DMPlex interfaces. This ensures that
computational models built using the Firedrake framework can easily be
extended without breaking existing abstractions and thus enables
domain scientists to leverage automated performance optimisations as
well as a wide range of simulation features.

\section{Acknowledgements}

This work was supported by the embedded CSE programme of the ARCHER UK
National Supercomputing Service \\(http://www.archer.ac.uk) and the
Intel Parallel Computing Center program through grants to both the
University of Chicago and Imperial College London. ML and GJG also
acknowledge support from EPSRC grants EP/M019721/1 and EP/L000407/1,
LM acknowledges NERC grant NE/K006789/1 and MGK acknowledges partial
support from DOE Contract DE-AC02-06CH11357 and NSF Grant OCI-1147680.

\bibliographystyle{siam}
\bibliography{references}

\end{document}